# Atomically precise step grids for the engineering of helical states

J. Enrique Ortega[1,2,3*], Guillaume Vasseur[3], Ignacio Piquero-Zulaica[2], Frederik Schiller[2], Julien Raoult[4], Miguel Angel Valbuena[5], Stefano Schirone[5], Sonia Matencio[5], Aitor Mugarza[5,6*], and Jorge Lobo-Checa[7,8]

[1]*Departamento Física Aplicada I, Universidad del País Vasco, 20018-San Sebastian, Spain*

[2]*Centro de Física de Materiales CSIC/UPV-EHU-Materials Physics Center, Manuel Lardizabal 5, 20018-San Sebastian, Spain*

[3]*Donostia International Physics Centre, Paseo Manuel de Lardizabal 4, 20018-San Sebastian, Spain*

[4] *Synchrotron SOLEIL, L'Orme des Merisiers Saint-Aubin, BP 48 F-91192 Gif-sur-Yvette Cedex, France*

[5]*Catalan Institute of Nanoscience and Nanotechnology (ICN2), CSIC and The Barcelona Institute of Science and Technology, Campus UAB, Bellaterra, 08193 Barcelona, Spain.*

[6]*ICREA Institució Catalana de Recerca i Estudis Avançats, Lluis Companys 23, 08010 Barcelona, Spain*

[7]*Instituto de Ciencia de Materiales de Aragón (ICMA), CSIC-Universidad de Zaragoza, E-50009 Zaragoza, Spain*

[8]*Departamento de Física de la Materia Condensada, Universidad de Zaragoza, E-50009 Zaragoza, Spain*

**Conventional spin-degenerated surface electrons have been effectively manipulated by using organic** [1,2] **and inorganic**[3] **self-assembled nanoarrays as resonators. Step superlattices naturally assembled in vicinal surfaces are a particularly interesting case since they represent simple one-dimensional (1D) models for fundamental studies, and can imprint strong anisotropies in surface electron transport in real devices. Here we present the first realization of periodic resonator arrays on the BiAg$_2$ atom-thick surface alloy with unprecedented atomic precision, and demonstrate their potential ability for tuning helical Rashba states. By employing curved crystals to select local vicinal planes we achieve tunable arrays of monoatomic steps with different morphology and orientation. Scanning the ultraviolet light beam on the curved surface during angle-resolved photoemission experiments allows one to unveil the scattering** behavior of spin-textured helical states. In this way, we find coherent scattering of helical Rashba states from the step arrays, as well as step-density-dependent Rashba band shifts and spin-orbit splitting compared to the extended BiAg$_2$ plane.

**Introduction**

Materials exhibiting Rashba spin-orbit coupling (SOC) have triggered an immense research activity that led to the opening of a new field in spintronics, the so-called spin-orbitronics.[4,5] As for any other material in the past, their successful technological application depends on two factors: a fundamental understanding of the emerging phenomena, and the capability to engineer their key properties. Exposing electrons with strong SOC to superlattice potentials is a very attractive approach for these two purposes. Indeed, different methods have been proposed for manipulating Rashba electrons by the interaction with one-dimensional (1D) periodic potentials. For instance, Rashba superlattices can induce standing spin waves[6] that alter the electric field induced spin accumulation only along the nanostructured periodic direction.[7] This characteristic spin texture can be further manipulated by using terahertz radiation.[8] The plasmonic response to such superlattices allows for a similar anisotropic tunability.[9] Recently, the use of periodic potentials has been proposed as scaffold to engineer the topology of Rashba superconductors aiming at the generation of Majorana Fermions.[10]

The fundamental study of Rashba electron's scattering can be unravelled when using 1D self-assembled superlattices. In conventional metal[2,11] and semiconductor[12,13] surfaces, such step arrays have been successfully employed as coupled resonator systems capable of tuning the overall electron confinement,[3,14] as well as for finding the coupling between the electronic and elastic energies of these nanostructures.[15,16] Adding the characteristic Rashba-type spin-momentum locking to surface electrons, the emergence of new scattering mechanisms can be expected, which in the presence of 1D periodic structures could give rise to exotic band renormalization, in a similar way to the one observed for pseudospin-momentum locked electrons in graphene.[17] However, the only Rashba superlattice studied to date consisted on the Shockley surface state on a stepped Au(111) surface, where the spin texture and polarization remains practically unaltered by the weak SOC.[18] More intriguing phenomena has been observed in individual step resonators fabricated on the basis of the BiAg$_2$ surface alloy grown on Ag(111).[19] In such system, the SOC is among the largest reported and its complex spin-orbital texture combined with the chemical heterogeneity has been shown to give rise to strongly anisotropic spin-flip scattering mechanisms.[20] However, the effect of a periodic potential on the Rashba electron gas of BiAg$_2$ or any other strong SOC system is still unexplored, since the realization of atomically precise step arrays in such materials remains a challenge.

Here we demonstrate that the fabrication of atomically-sharp BiAg$_2$ step superlattices with variable periodicity is feasible. Depositing 1/3 ML of Bi on curved vicinal Ag(111) surfaces, we obtain BiAg$_2$ surface alloys that replicate the step array of the vicinal template. The different step morphologies can be experimentally accessed by selecting the azimuthal angle that defines the orientation of the cylindrical section of the curved crystal, and hence the step direction and atomic structure (see Fig. 1**a**). Following scanning tunneling microscopy (STM) imaging and low energy electron diffraction (LEED), we find particularly stable, straight step superlattice spaced with atomic precision at specific azimuthal directions, revealing the stabilizing role of the basic surface alloy structure. Moreover, two local planes along the cylinder match integer number of the BiAg$_2$ alloy unit cell, which result on unimodal, atomically sharp superlattices with unprecedented qualities. We investigate the Rashba bands on such "magic" planes via angle resolved photoemission (ARPES)[11,15,16,21]. We observe coherent scattering of helical Rashba states from the step arrays, which result in step-density-dependent Rashba band shifts and variable SOC-splitting with respect to the extended BiAg$_2$(111) plane.

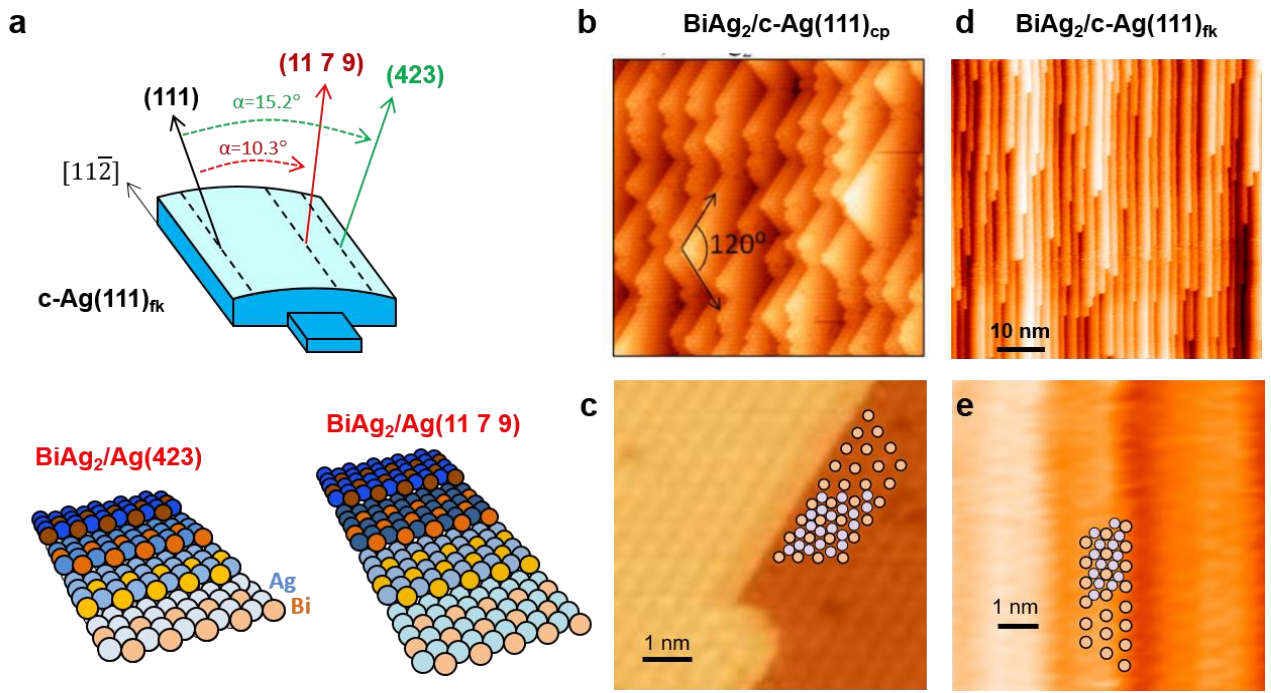

**Figure 1.** Morphology of the BiAg$_2$ surface alloy on vicinal Ag(111) surfaces with different step types. **a** Sketch of the c-Ag(111)$_{fk}$ curved crystal sample used to systematically investigate BiAg$_2$ growth at fully kinked Ag(111) vicinal surfaces. Magic Ag(11 7 9) and Ag(423) planes are tilted $\alpha=10.3°$ and $\alpha=15.2°$ away from the (111) plane, respectively. **b-c** STM images of stoichiometric BiAg$_2$ on the c-Ag(111)$_{cp}$ curved sample featuring closed-packed steps. Bi-termination of step-edges drives the zigzagging into 120°-rotated fully kinked step segments. The STM tip probes the Bi sublattice. **d-e** STM images of stoichiometric BiAg$_2$ on the c-Ag(111)$_{fk}$ curved sample with fully kinked steps. Bi termination of step-edges in this case favors long, defect-free straight steps.

**Results**

Room temperature deposition of 0.33 monolayers (ML) Bi followed by a mild post-annealing (550 K) on the curved Ag(111) surface featuring close-packed steps [c-Ag(111)$_{cp}$ sample], leads to perfect BiAg$_2$ surface stoichiometry at all vicinal planes, but does not result in the desired structural ordering of the step lattice, as shown in Fig. 1**b-c** (see Fig. S1 for a complete set of STM pictures). On the terraces one can resolve the characteristic (√3×√3)R30° atomic structure of the Bi sub-lattice, but steps reconstruct in a 120° zigzagged geometry (cf. Fig. 1**b**), that is, forming alternate segments of fully kinked steps along [11-2] and [-2-11]. As schematically overlaid using the atomic model of Fig. 1**c**, the zigzag roughening points to the presence of Bi-terminated step edges,[16] which, for the BiAg$_2$ stoichiometry, is only possible at fully kinked steps. Such Bi "segregation" at steps is analog to the 2D surfactant effect observed during Ag growth on Bi/Ag(111),[22] which is explained by the lower surface free energy of Bi ($\gamma_{0,Bi}$ = 0.53 J/m$^2$) compared to that of Ag ($\gamma_{0,Ag}$ = 1,2 J/m$^2$). To avoid zigzagging and favor vicinal BiAg$_2$ surfaces with parallel arrays we use the c-Ag(111)$_{fk}$ sample with fully kinked steps oriented along the [11-2] direction. As shown in Fig. 1**d-e**, the same 0.33 ML Bi evaporation and annealing process on c-Ag(111)$_{fk}$ leads to a radically different morphology, with stoichiometric BiAg$_2$ terraces and perfectly straight, almost defect-free monatomic ($h$=2.36 Å) steps.

We carry out a systematic structural exploration of the BiAg$_2$ monolayer across the c-Ag(111)$_{fk}$ template using LEED and STM. Top panels in Figs. 2**a** and 2**b** show selected LEED patterns acquired at different vicinal angles before and after BiAg$_2$ alloy formation. The excellent BiAg$_2$ lattice ordering is demonstrated by the appearance of sharp (√3×√3)R30° spots, whereas the persistence of the step lattice splitting proves that a regular monatomic step spacing is kept at all vicinal planes. The step-lattice or terrace-reconstruction nature of the diffracted beams is readily recognized in the LEED scans shown in the corresponding bottom panels. Each image is constructed with individual profiles along the line joining the (-1,0) and (-⅓,-⅓) spots (delimited by the red-dashed boxes above) from patterns acquired at 36 different vicinal angles ($\alpha$). The clean substrate image presents two characteristic straight lines that converge at the (111) plane, as expected for a linear variation of the step density $1/d$, being $d=h/\sin\alpha$ the step spacing or terrace size.[16] Notably, for the BiAg$_2$ alloyed surface the crossed lines appear replicated, supporting that the step array structural perfection is generally preserved. Moreover, at the local plane (miscut) angles of $\alpha$~10° and $\alpha$~15°, that is, in the vicinity of the (11 7 9) and (423) planes, the step-related periodicity and the reconstruction pattern nest. In this way, the LEED spots exhibit defined sharp rectangular and oblique patterns, defining two vicinal "magic" arrangements, as schematically shown in Fig. 1**a** bottom.

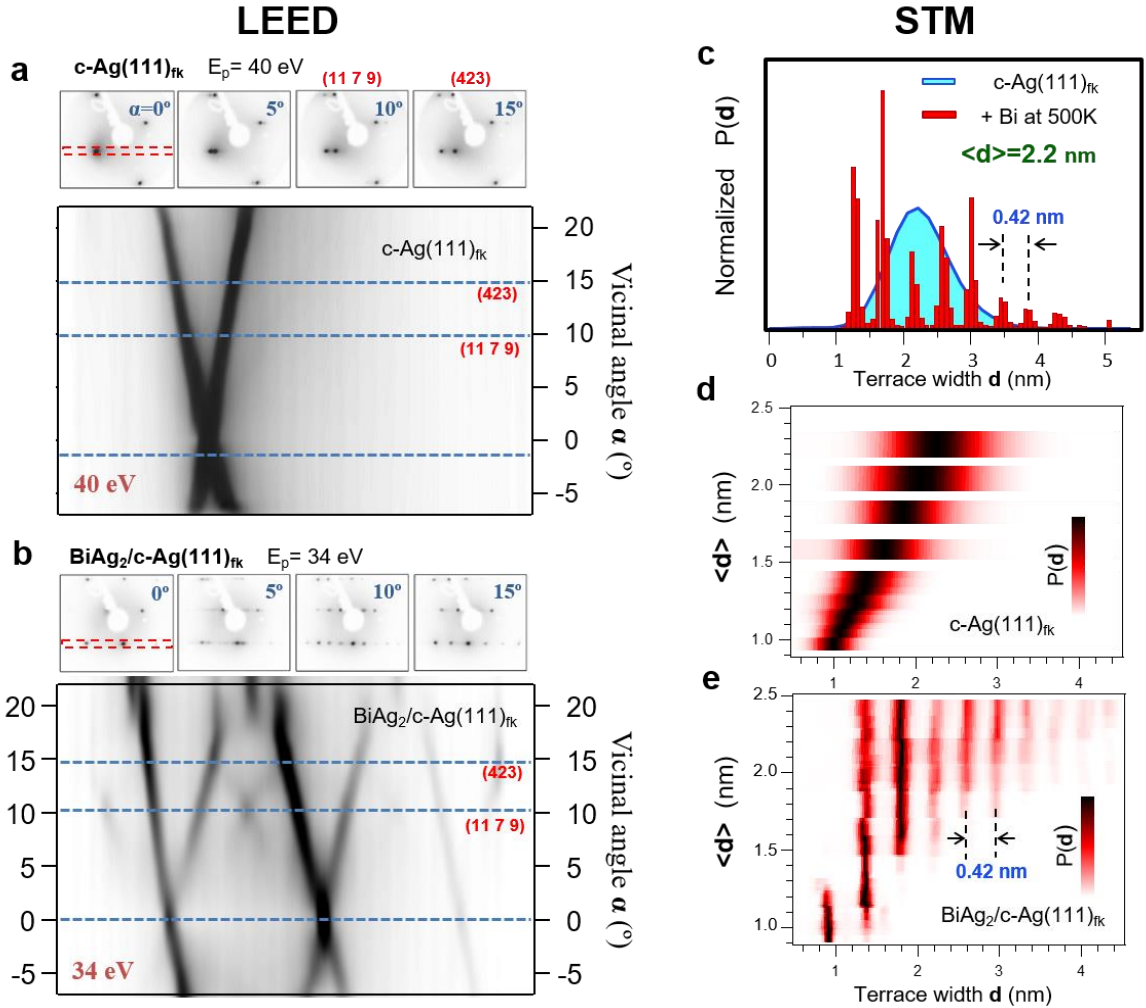

**Figure 2.** Structural analysis of the fully kinked curved crystal before and after BiAg$_2$ alloy formation. **a** LEED patterns at selected $\alpha$ vicinal angles on the clean c-Ag(111)$_{fk}$ sample. At $\alpha$=10º and $\alpha$=15º we probe the magic (11 7 9) and (423) planes, respectively. The bottom panel shows the LEED image constructed with individual profiles along the (-1,0), marked by the red rectangle above. The straight crossing lines reflect the smoothly varying step density of the curved surface. **b** Identical LEED analysis corresponding to the BiAg$_2$/c-Ag(111)$_{fk}$ system. The sharp ($\sqrt{3}\times\sqrt{3}$)R30° pattern indicates a perfect BiAg$_2$ stoichiometry in all vicinal planes. At magic (11 7 9) and (423) local planes, a perfect step/reconstruction matching leads to extremely sharp LEED spots. Once again, the bottom panel corresponds to the LEED image constructed with individual profiles along the (-⅓,-⅓), marked by the red rectangle above. **c** Terrace width distribution (TWD) analysis of STM images for 2.2 nm terraces of the pristine (blue) and BiAg$_2$ alloy on the c-Ag(111)$_{fk}$ curved crystal. The smooth histogram of the former is broken into 0.42 nm "quanta" in the latter, as expected for perfect BiAg$_2$ stoichiometry and Bi-terminated step edges (see text for details). **d-e** TWD images across the clean and the BiAg$_2$-covered c-Ag(111)$_{fk}$ crystal. Upon Bi growth, 0.42 nm quantization arises at the full set of vicinal planes.

The structural quality of the BiAg$_2$ step lattices grown onto the c-Ag(111)$_{fk}$ crystal is further examined at the nanoscale level by means of STM. The local fluctuations of the step spacing $d$ around its average value $<d>$, caused by the thermal excitation of kink atoms at stepped substrates, provides a good measure of the sharpness of the array.[23,24] The terrace-width fluctuation of the BiAg$_2$-covered vicinal surface of Fig. 1**d** is represented in the probability plot of Fig. 2**c.** Using and automated line-by-line image analysis process[16], we obtain the statistical variation [**P**($d$)] of the terrace-width $d$ (also called terrace-width distribution (TWD)) represented by the bar histogram (in red). It can be compared with the **P**($d$) function of the bare vicinal plane (blue curve) measured at the same α=6° vicinal angle (or mean value $<d>$=2.2 nm).[16] After the formation of the alloy, the smooth **P**($d$) curve of the clean surface is broken up in a periodic series of sharply defined maxima, with period equal to the Bi row spacing **a$_0$**=0.42 nm of the (√3×√3)R30° reconstruction (**a$_0$**=√3**a$_{Ag}$**).

The radical step array transformation into defined bunches or **a$_0$** "quanta" is again forced by the Bi termination of step edges in a perfect BiAg$_2$ stoichiometry. The phenomenon, which is analogous to the one observed in vicinal Si(111) surfaces, where the 7×7 reconstruction pattern is the driving force of terrace quantization and sharp termination of step-edges[26,27], takes place at all α angles. In Fig. 2**d-e** we show in a color scale the respective **P**($d$) functions across the clean and the BiAg$_2$-covered c-Ag(111)$_{fk}$ surface. At the clean sample (Fig. 2**d**), one can readily observe the characteristic α-dependence of the **P**($d$) probability function of vicinal metal surfaces,[16,25] that is, the smooth transition from asymmetric-and-broad **P**($d$) at small α angles, to symmetric-and-sharp **P**($d$) at larger α. Upon BiAg$_2$ alloy formation (Fig. 2**e**), the **P**($d$) plot histogram transforms in a set of vertical streaks indicating that **a$_0$** quantization occurs at all vicinal planes along the curved surface. The terrace-width distribution map of Fig. 2**e** demonstrates the big potential of the curved surface approach for a proper identification of optimal substrates where sharp BiAg$_2$ step arrays are achieved. The **P**($d$) histogram notably sharpens below $<d>$ ~1.5 nm (α=9°). Careful inspection pinpoints two vicinal planes at which the histogram presents a single **a$_0$** peak, at $d$=1.36 nm (α~10°) and $d$=0.93 nm (α~15°). These particular local planes correspond to "magic" angles with *atomically sharp* step lattices, stabilized by the surface alloy structure.

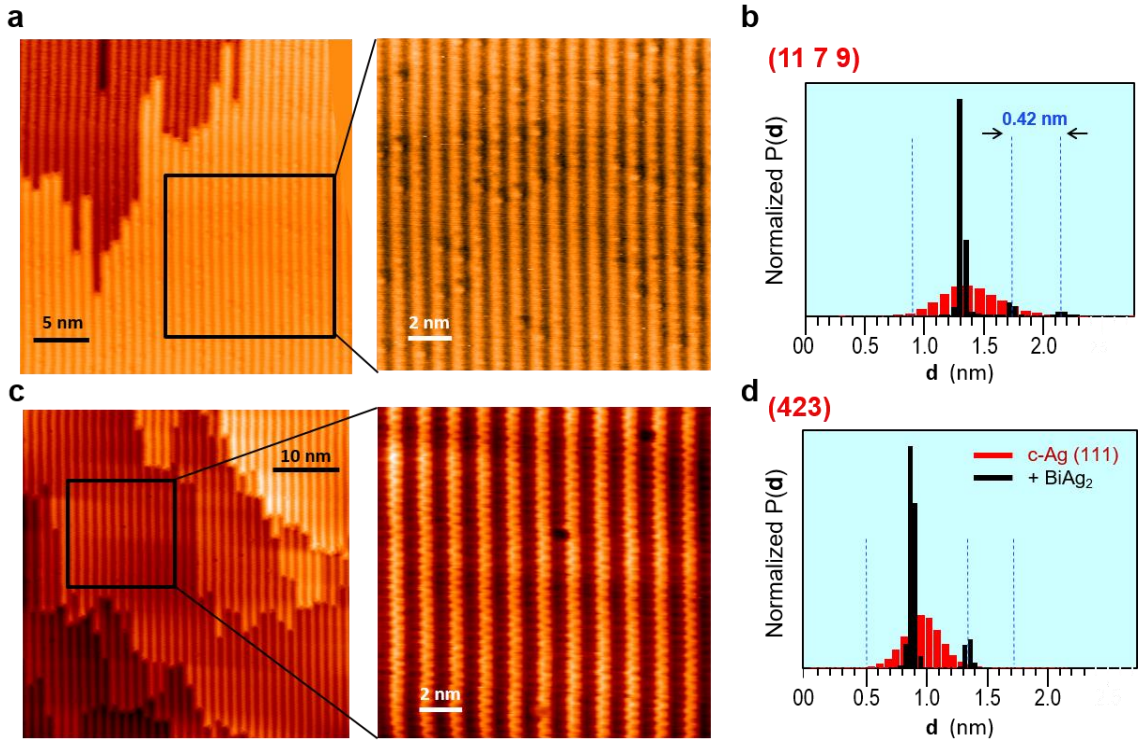

**Figure 3.** STM topographic images from BiAg$_2$ surface alloy at the local **a** Ag(11 7 9) and **c** Ag(423) planes on the c-Ag(111)$_{fk}$ sample. Both positions exhibit sharp, almost defect-free monatomic step arrays. **b-d** Terrace-width distribution (TWD) derived from statistical STM image analysis at BiAg$_2$/Ag(11 7 9) and BiAg$_2$/Ag(423) surfaces. The resulting surface alloy histograms (black) are compared with the TWD of their respective clean substrates (red in the back). BiAg$_2$ growth leads to the almost complete elimination of terrace size fluctuations around the expected $d$ value of the ideal Ag(11 7 9) and Ag(423) crystal truncation planes.

In Fig. 3 we show the STM images taken at these two selected positions on the curved surface, together with their corresponding histograms. These reveal BiAg$_2$ step lattice grids of atomic-scale quality (black bars in the histogram), with residual presence of other terrace sizes following the absence of counts at contiguous **a$_0$** quanta (marked by the adjacent vertical dotted lines). The relatively sharp **P**($d$) histograms ($\sigma = \Delta d/d$ ~0.3) for the corresponding clean vicinal surfaces are shown in red in the background. Their shape appear as envelopes of the BiAg$_2$ histograms, forcing possible terrace sizes to a single **a$_0$** quantum, therefore preventing the presence of adjacent quanta. Since the size dispersion $\sigma$ in the clean Ag vicinal substrate is determined by the step-step repulsion,[23,24] our quantitative analysis for the alloyed surface "magic" angle planes suggests a dominating substrate-step interaction. Such atomically-sharp (single **a$_0$** quantum) BiAg$_2$ step grids can be achieved below a critical $d <$ **a$_0$**/$\sigma$ ~1.4 nm step spacing, as is shown in Fig. 2**e**.

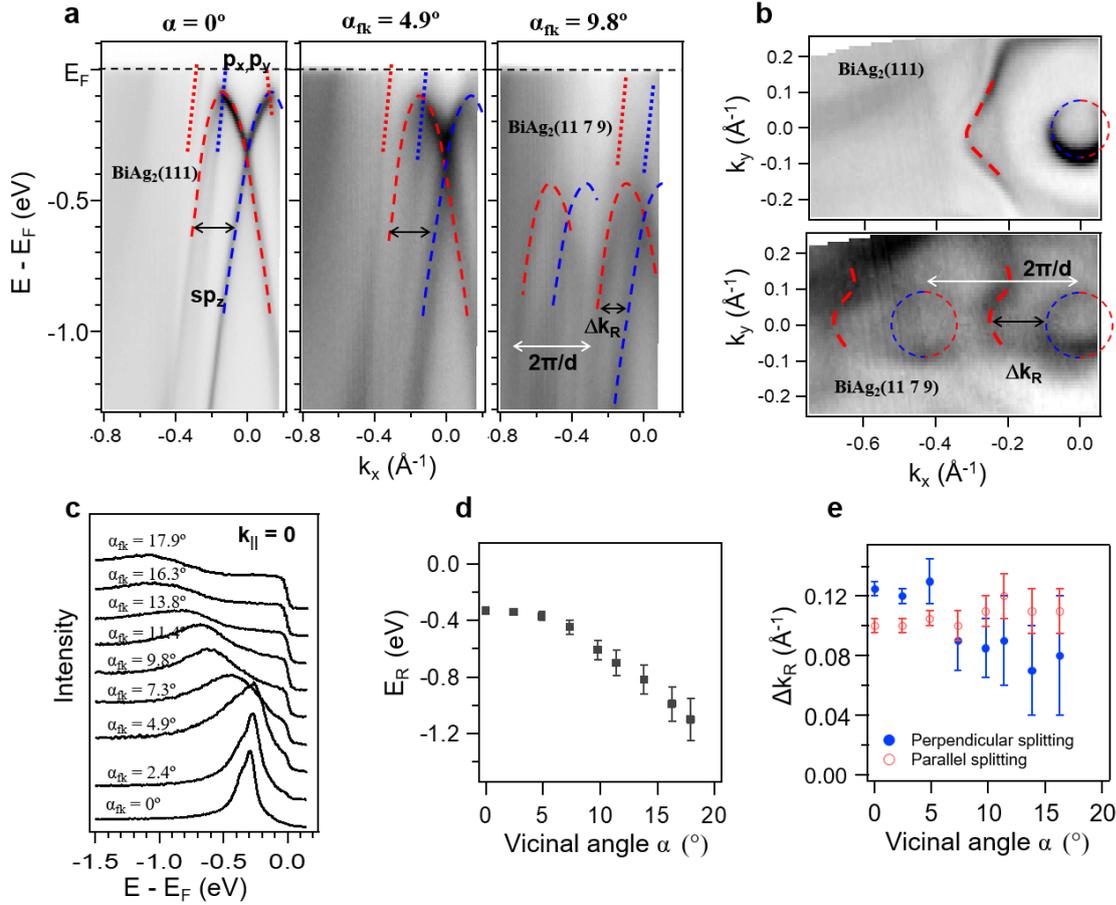

**Figure 4. a** Rashba helical states (red and blue mark different helicities) measured at three characteristic points on the BiAg$_2$/c-Ag(111)$_{fk}$ system. The panel close to 10° corresponds to the (11 7 9) surface shown in Fig. 3**a**. 2π/d umklaps are identified through a careful inspection of second derivative plots (see S.I.), and constant energy surfaces across the curved surface. **b** Constant energy surfaces at BiAg$_2$(111) and BiAg$_2$(11 7 9). **c** Energy distribution curves at k=0 for different local plane (miscut) angles. The peak corresponds to the Rashba point, which is plotted in **d** as energy vs vicinal angle. **e** Rashba splitting parallel and perpendicular to the step array, as a function of the vicinal angle.

The excellent quality of the BiAg$_2$ step array grids on the c-Ag(111)$_{fk}$ should lead to well-defined, sharp Rashba states in ARPES measurements. Indeed, Fig. 4 shows excellent data quality of the spin-split bands exhibiting the characteristic signatures of Bragg scattering at step superlattices,[11,15,16] i.e. umklapp bands and size effects. In Fig. 4**a-b** we show at selected points on the curved BiAg$_2$ surface the band dispersion and constant energy surfaces (CES-s) where the Rashba states are dominant (see SI for more surface bands and CES-s). At the (111) plane position we observe the characteristic Rashba band manifold of BiAg$_2$, with the upper lying $p_x$-$p_y$ bands crossing E$_F$, and the prominent pair of $sp_z$ Rahsba helical states with apex (band maximum) at 0.1 eV.[19] When the beam is focused at the vicinal angle of α~5°, a similar band structure to the (111) is observed accompanied with significant spectral broadening. This α~5° position (mean value <d>~2.7 nm) corresponds to a BiAg$_2$ surface with multiple $a_0$ terrace-quanta (cf. Fig. 2e), such that no coherent scattering by the step lattice should be expected[1]. Coherent step lattice scattering emerges at the sharp BiAg$_2$(11 7 9) surface near the magic $α=10°$ vicinal angle, leading to radical changes in the $sp_z$ Rashba bands. Although the weak intensity and the interference arising from the overlap with $p_x$-$p_y$ bands complicates their respective identification, a joint inspection of direct and second derivative images and CES-s (see Figs. S3 and S4) allows us to label the helical states pair and corresponding step-lattice umklapps. The Rashba point remains at $k_x=0$, but exhibits a strong shift to higher binding energy and a slight reduction of the Rashba splitting at high vicinal angles, as shown through Figs. 4c-e. The downward shift of the Rashba bands may simply reflect that steps act as repulsive potential barriers for these helical hole-like states. As quantitatively determined for Shockley states in curved noble metal surfaces,[15] the step barrier strength can be estimated from the systematic downward shift measured across the curved surface, assuming a simple 1D Kronig-Penney model for the step lattice (see Fig. S3). This renders a barrier strength of $V_0 \times b =$ 10 eV.Å ($V_0$, barrier height, b, barrier width) that is much higher than step barriers found in bare vicinal Ag(111) surfaces with fully kinked steps (see Fig. S2).[16]

In summary, we have shown that by using the curved crystal approach we can tune the growth of a BiAg$_2$ monolayer alloy featuring sharp, Bi-terminated step of varied density. Moreover, we can achieve arrays exhibiting "magic" monoatomic step periodicities whenever integer numbers of the BiAg$_2$ unit cell can be fit within small average terrace sizes (below $<d>$ ~1.4 nm or $α > 9°$). Furthermore, ARPES experiments allows us to unveil coherent scattering in the spin-textured helical Rashba states, characterized by step-density-dependent energy shifts and SOC-splitting.

**Methods**

The curved crystal surface of Ag were prepared by repeated cycles of sputtering, with Ar$^+$ at an energy of 1 KeV and incident angle of 45° in the direction parallel to steps, and mild annealing to 700 K for 15 min at a maximum pressure of $5 \times 10^{-10}$ millibar. Both c-Ag(111)$_{fk}$ and c-Ag(111)$_{cp}$ were fabricated (Bihurcrystal Ltd, Spain) by mechanical erosion and polishing of (645)- and a (111)-oriented Ag wafers, such as to define 24° and 16° cylindrical sections, with axis parallel to the [11-2] and [1-10] crystal directions, respectively. The c-Ag(111)$_{fk}$ sample was mainly used in the present work, whereas the c-Ag(111)$_{cp}$ sample, featured with variable densities of A-type ({100}-like minifacets) and B-type ({111}-like minifacets) close-packed steps was used for comparative STM experiments (see Fig. S1).

---

[1] As observed in Shockley states of vicinal noble metal surfaces (10), the lack of coherence in step lattices decouples terraces and leads to a broader (111)-like band. Rashba state broadening is even larger for the BiAg$_2$ layer grown on the c-Ag(111)$_{cp}$ surface, due to the general disorder of the step lattice shown in Fig. 1**b-c.**

Bi was evaporated from a Knudsen cell with a rate 0.06 ML/min with the sample held at 300 K, and then a gentle post-annealing at 550 K results in the formation of the BiAg$_2$ monolayer alloy. For STM and LEED experiments we ensure the accurate BiAg$_2$ stoichiometry by evaporating, at 300 K, a shallow 0.2-0.4 monolayer wedge in the direction parallel to the steps, and then selecting the precise 0.33 Bi ML position on the sample, as judged from LEED. For ARPES experiments a quartz-microbalance calibration is previously obtained to achieve a direct 0.33 ML evaporation for the experiment. The quality of the surface alloy is further checked by judging the sharpness of the spin-orbit split bands at the (111) plane.

LEED measurements were conducted at 300 K, using a standard 3-grid optics system (Omicron). Specific points on the c-Ag(111)$_{fk}$ curved sample were addressed with the minimum 300μm spot size, which spreads over Δα~0.35° arc on the curved surface. Such spreading of the electron beam explains the spot-splitting broadening observed in LEED images from the bare c-Ag(111)$_{fk}$ (see Fig. 2**a**).

STM imaging was carried out using a variable temperature setup (Omicron) operating at 300 K. The tunneling current was set to ~0.1 nA, and the sample bias either to -1.0 V or +1.1 V. To elaborate the probability **P**(*d*) histograms, STM images were acquired at the selected surface areas exhibiting homogeneous step arrays in the μm scale. Topography data were analyzed using a home-programmed code that automatically finds step edges and delivers the **P**(*d*) histogram for each image.[16] **P**(*d*) data shown in Figs. 2 and 3 correspond either to a single or to the average of two contiguous STM frames, where sizes vary between 40×40 and 300×300 nm$^2$, in order to image between 20 and 40 steps within the same frame.

ARPES measurements were performed at the CASSIOPEE beamline of SOLEIL Synchrotron (Saint-Aubin, France). We used linearly polarized photons of 21 eV and a hemispherical electron analyzer with vertical slits with +-15° angular acceptance, which provides high-resolution band mapping by moving the polar angle of the manipulator. The photon beam impinged the sample with a 45° angle.

The spot size on the sample was ~50×15 μm$^2$, with the wider axis aligned parallel to the steps. The 15 μm length perpendicular to the step direction defined a negligible 0.02° spread along the curve of the crystal, that is, a sharp sensitivity to the local crystallographic orientation. The ARPES experiments were carried out at low temperature (100 K) and the angle and energy resolutions were set to 0.05° and 10 meV, respectively.

**Acknowledgements**

This research was supported by the CERCA Programme/Generalitat de Catalunya, and funded by the Spanish Ministry of Science, Innovation and Universities (Grants MAT2016-78293-C6 (2-R, 4-R and 6-R), MAT-2017-88374-P and Severo Ochoa No. SEV-2017-0706), the Basque Government (Grant IT-1255-19), the regional Government of Aragón (RASMIA project) and the European Regional Development Fund (ERDF) under the program Interreg V-A España-Francia-Andorra (Contract No. EFA 194/16 TNSI).

# Supplementary Information

# Atomically precise step grids for the engineering of helical states

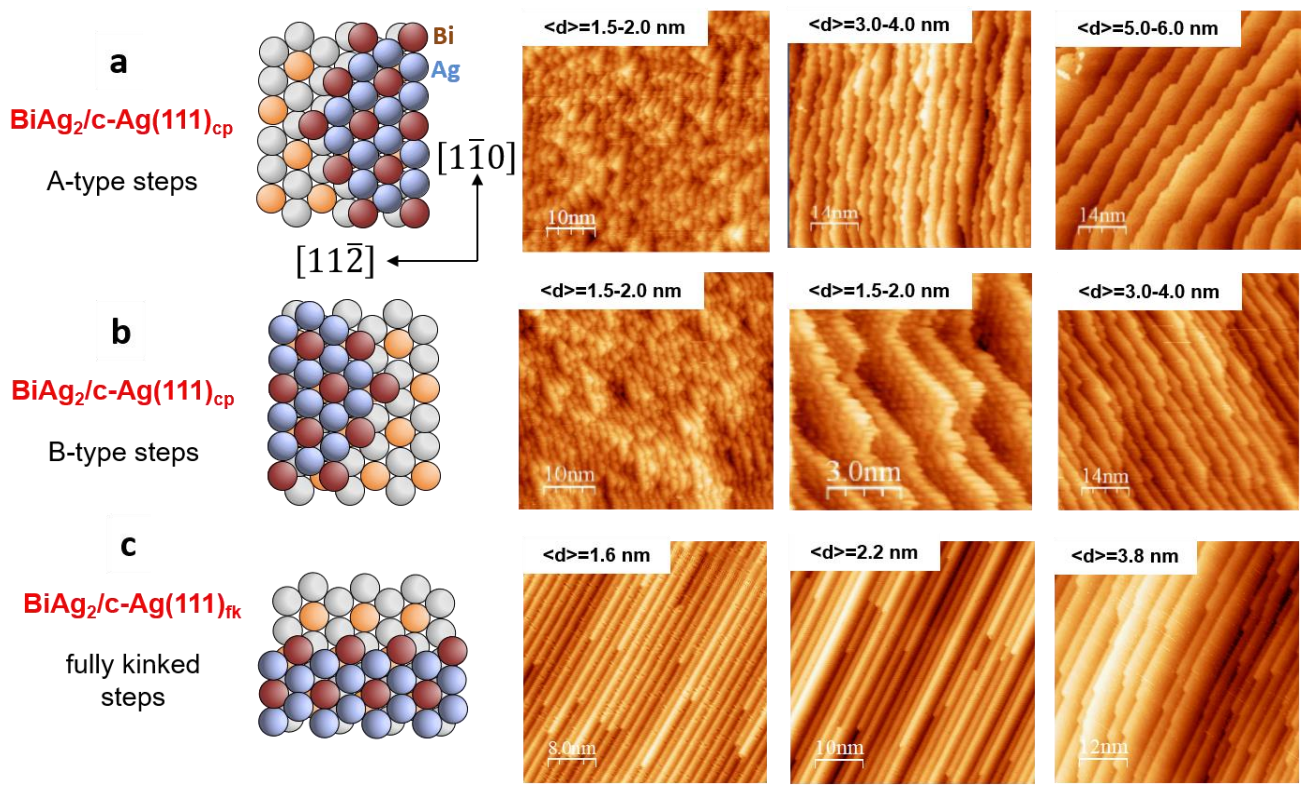

**Figure S1.** Left, schematic description of the atomic arrangement at the step edge, and right, selected STM topographies corresponding to different vicinal angles (<*d*> average terrace values) of BiAg$_2$-covered vicinal surfaces with **(a)** A-type close-packed steps, **(b)** B-type close-packed steps, and **(c)** fully kinked steps. Vicinal surfaces with close-packed A- and B-type steps were examined using a cylindrical c-Ag(111)$_{cp}$ sample, with cylinder axis parallel to the [1-10] direction (see Ref. [16] in the main text). Bi-termination of step-edges forces zigzagging along fully kinked directions, equivalent to the [11-2] crystal orientation. For surfaces with steps oriented along [11-2] (fully kinked vicinals) in **(c),** Bi-termination of step-edges leads to straight, defect-free steps at all vicinal angles.

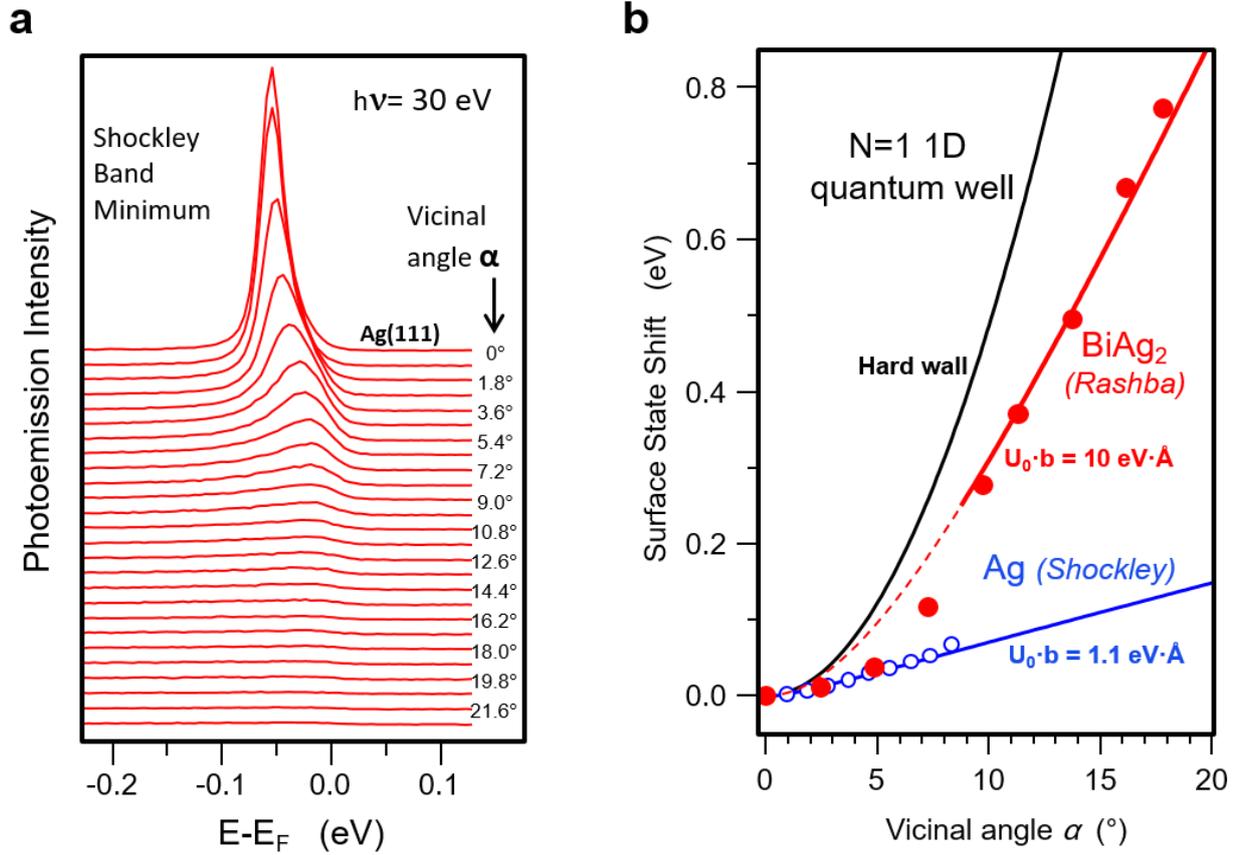

**Figure S2. (a)** Shockley state peak measured at band minimum ($k_x=\pi/d$, see Refs. [11 and 16] in the main text) across the c-Ag(111)$_{fk}$ sample, exhibiting the expected upward shift as a function of the miscut angle, which reflects the size effect for surface electron confinement within (111) terraces (also called terrace-size effect, see Ref. [16] in the main text). **(b)** Terrace-size effect for the Shockley state of clean (blue) and for the Rashba-split state of the BiAg$_2$-covered (red) fully kinked vicinals (see also Fig. **4d** of the main text), as measured using the c-Ag(111)$_{fk}$ sample. Lines are fits to a 1D Kronig-Penney (KP) model for the periodic step potential, from which a step barrier strength $U_0 \cdot b$ (see Refs. [11,15, 16] in the main text for details) is obtained. The black solid line represents the energy shift for the N=1 state in the 1D infinite potential well of width $d$ [$\Delta E = (\hbar^2\pi^2/2m^*d^2)$]. Below $\alpha=10°$, the lack of a truly coherent step lattice limits the validity of the model, whereas above $\alpha=10°$ the KP model fits reasonably well. However, the large barrier strength obtained ($U_0 \cdot b$ ~10 eV·Å) is inconsistent with the highly dispersive Rashba-split bands observed (Fig. **4a** of the main text and Fig. S3), indicating that the simple 1D KP model breaks down.

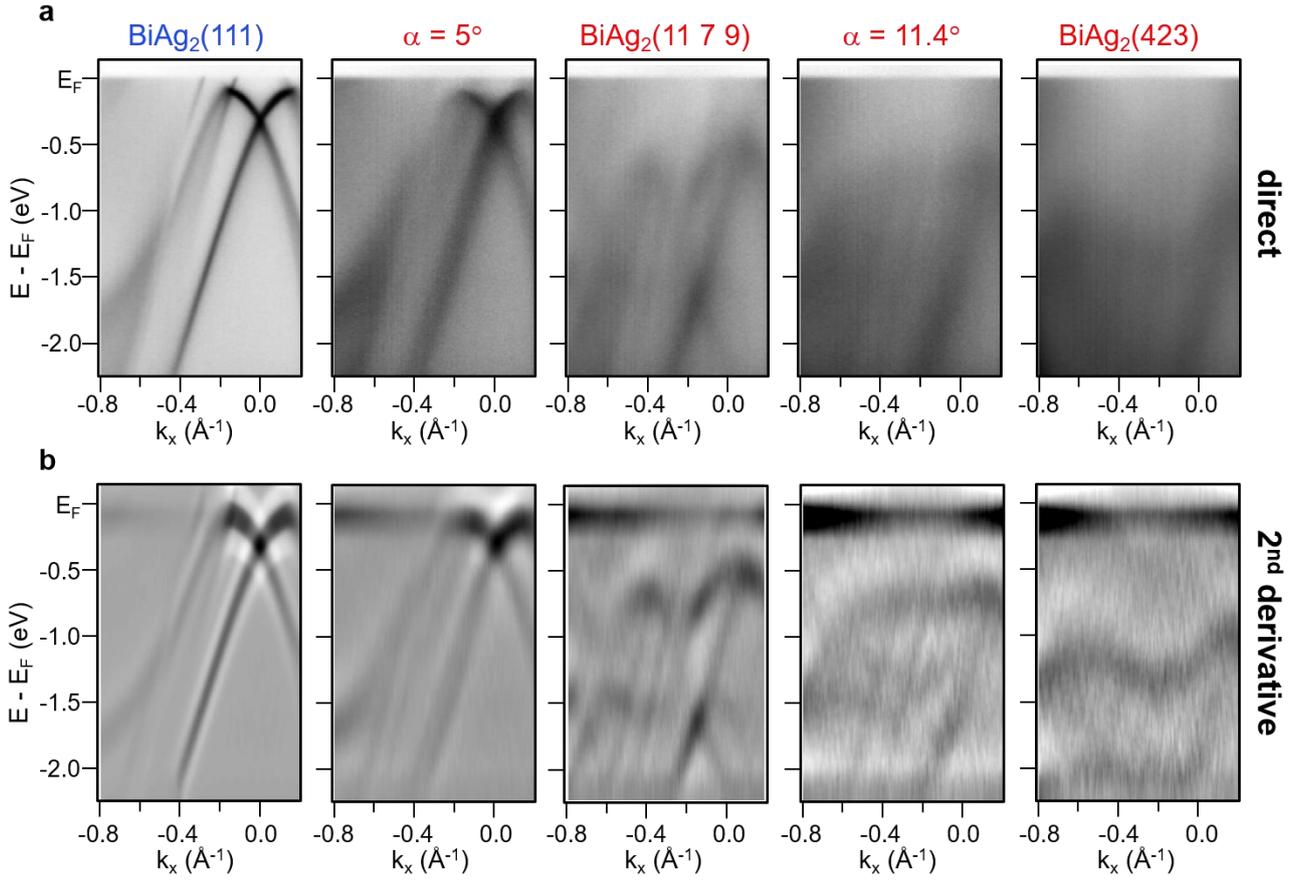

**Figure S3.** Rashba-split bands for selected cases across the BiAg$_2$-covered c-Ag(111)$_{fk}$ sample. The photoemission intensities are shown in **(a)** and their corresponding second derivatives in **(b)** as gray scale plots, for the angles indicated on top. The wave vector scale refers to the electron momentum with respect to the (111) surface plane and follows the direction perpendicular to the steps. The photon energy is set to 21 eV.

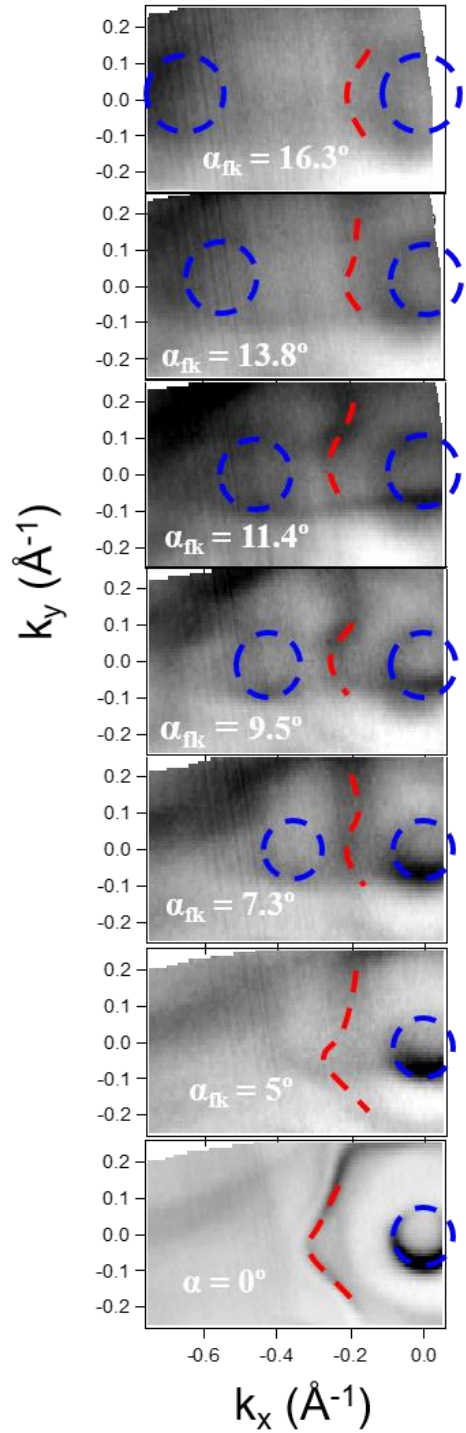

**Figure S4.** Fermi surface measured as a function of the vicinal angle $\alpha_{fk}$ on the BiAg$_2$-covered c-Ag(111)$_{fk}$ sample, shown in a gray scale that is proportional to the photoemission intensity. The wave vector scale refers to the electron momentum with respect to the local surface plane, with $k_x$ and $k_y$ perpendicular and parallel to the kinked step direction, respectively. The photon energy is set to 21 eV. Lines are guide to the eyes, and refer to the average clockwise (red) and counterclockwise (blue) spin rotations on the surface plane.

17